\documentclass[%
 reprint,
 amsmath,amssymb,
 aps,
 prd
]{revtex4-2}

\usepackage[utf8]{inputenc}
\usepackage{tensor}
\usepackage{hyperref}
\usepackage{cleveref}
\usepackage{graphicx}
\usepackage{bm}
\usepackage{xcolor}

\usepackage{cleveref}
\crefname{equation}{Eq.}{Eqs.}
\crefname{appendix}{Appendix}{Appendices}
\crefname{figure}{Fig.}{Figs.}

\begin{document}

\title{Observation of gravitational waves by light polarization}

\author{Chan Park}
    \affiliation{National Institute for Mathematical Sciences, Daejeon 34047, Republic of Korea}
    \email{iamparkchan@gmail.com}

\author{Dong-Hoon Kim}
    \affiliation{Department of Physics and Astronomy, Seoul National University,
Seoul 08826, Republic of Korea}
    \email{ki13130@gmail.com; corresponding author}

\date{\today}

\begin{abstract}
We provide analysis to determine the effects of gravitational waves on
electromagnetic waves, using perturbation theory in general relativity. Our
analysis is performed in a completely covariant manner without invoking any
coordinates. For a given observer, using the geometrical-optics approach, we
work out the perturbations of the phase, amplitude, frequency and
polarization properties--axes of ellipse and ellipticity of light, due to
gravitational waves. With regard to the observation of gravitational waves,
we discuss the measurement of Stokes parameters, through which the antenna
patterns are presented to show the detectability of the gravitational wave
signals.
\end{abstract}

\maketitle

\section{Introduction}

Light serves as a powerful means to observe gravitational waves. In fact,
most of the current and future gravitational-wave detectors make use of
light. As a primary example, laser interferometers observe gravitational
waves through the interference patterns of light caused by the difference of
the photon transit times between two arms \cite%
{malik_rakhmanov_response_2007,rakhmanov_high-frequency_2008,rakhmanov_round-trip_2009}%
. Viewing this from a different perspective, the electromagnetic waves
propagating along each arm are perturbed by gravitational waves, and the
observation of gravitational waves is enabled by the difference between the
phases of the perturbed electromagnetic waves from the two arms: the
electromagnetic waves interfere when they meet at the intersection of the
arms, where a trace of the gravitational wave signals would cause changes to
the intensity of the superposed electromagnetic waves \cite{thorne2017}. As
another example, pulsar timing arrays observe gravitational waves through
the changes in the electromagnetic pulse periods from pulsars. The phase of
electromagnetic waves is delayed during the passage of gravitational waves,
which eventually causes the measured pulse frequency (or period) to vary
slightly. Then the cumulative variation of this (termed as a \textit{residual%
}) enables the observation of gravitational waves \cite%
{detweiler_pulsar_1979}: taking account of the cross-correlation of the
residuals of two pulsars nearby in the sky (i.e., the quadrupolar
interpulsar correlation) would enhance confirmation of the observation \cite%
{hellings_upper_1983}.

There have been numerous studies investigating the effects of gravitational
waves on light in the context of general relativity. Some of the studies
focus on the effects on the polarization of light. Among others, Montanari 
\cite{montanari_propagation_1998} obtained a solution of Maxwell equations
in a gravitational wave background to one order of approximation beyond the 
\textit{geometrical-optics} limit, and applied this to the study of
perturbations of the linear polarization of electromagnetic waves. Calura
and Montanari \cite{calura_exact_1999} presented the exact solution to the
linearized Maxwell equations in spacetime slightly curved by a gravitational
wave only in the framework of the linearized general relativity, without
invoking the geometrical-optics approximation, and applied this to the case
of a linearly polarized electromagnetic field bounced between two parallel
conducting planes. Halilsoy and Gurtug \cite{halilsoy_search_2007} analyzed
the Faraday rotation in the polarization vector of a linearly polarized
electromagnetic shock wave upon encountering with gravitational waves.
Hacyan \cite{hacyan_electromagnetic_2012,hacyan_effects_2016} determined the
influence of a gravitational wave on the elliptic polarization of light,
deducing the rotation of the polarization angle and the corresponding Stokes
parameters, and applied this effect to the detection of gravitational waves,
as a complement to the pulsar timing method. Cabral and Lobo \cite%
{cabral_gravitational_2017} obtained electromagnetic field oscillations
induced by gravitational waves and found that these lead to the presence of
longitudinal modes and dynamical polarization patterns of electromagnetic
radiation.

In this paper, we employ perturbation theory of general relativity to
analyze the influence of gravitational waves on electromagnetic waves,
concentrating mainly on the effects on the polarization of light. Our
analysis is then applied to the observation of gravitational waves by means
of Stokes parameters. Largely, the paper proceeds in three steps through %
\Cref{sec:preliminaries,sec:perturbation_of_emw,sec:application}. In %
\Cref{sec:preliminaries}, we review the basics of gravitational and
electromagnetic waves as described in the flat spacetime background, and
introduce our notational conventions used in the paper. In %
\Cref{sec:perturbation_of_emw}, we work out a perturbation of
electromagnetic waves due to gravitational waves from the perturbed Maxwell
equations: using the geometrical-optics approach, the perturbations of the
phase, amplitude, frequency and polarization properties--axes of ellipse and
ellipticity of light are determined. In \Cref{sec:application}, application
of our analysis to the observation of gravitational waves is discussed.
Stokes parameters are employed as optical observables to identify the
gravitational wave signals from, and measured in a suitable observational
frame. The antenna patterns are defined via the Stokes parameters to exhibit
the detectability of the gravitational wave signals.

Throughout the paper, our analysis is conducted in a completely covariant
manner without invoking any coordinates. To express tensors, Roman indices ($%
a$, $b$, $c$, $...$) are used; however, they should be distinguished from
other unitalicized or Greek or parenthesized subscripts used occasionally
for special notations; e.g., $M_{\mathrm{o}}$, $M_{\epsilon }$, $\omega _{%
\mathrm{g}}$, $\omega _{\mathrm{e}}$, $\mathcal{E}_{\left( p\right) }$, $%
\mathcal{E}_{\left( s\right) }$, etc. Also, a parenthesized number on the
top left of the surrounding text denotes the degree of perturbation; e.g., $%
^{\left( 0\right) }T_{ab}$ for the unperturbed tensor $T_{ab}$, $^{\left(
1\right) }T_{ab}$ for the first-order perturbation of $T_{ab}$, etc. We use
the geometrized unit system ($c=1$, $G=1$) for gravitation and the Gaussian
unit system ($\epsilon _{0}=\frac{1}{4\pi }$, $\mu _{0}=4\pi $) for
electromagnetism.

\section{Preliminaries}

\label{sec:preliminaries}

To discuss a perturbation of a quantity $X$ in a spacetime\ $\left( M_{%
\mathrm{o}},\tensor[^{\left(0\right)}]{g}{}\right) $, where $M_{\mathrm{o}}$
denotes a background manifold with a metric $\tensor[^{\left(0\right)}]{g}{}$%
, we consider a one-parameter family of perturbed spacetimes $\left(
M_{\epsilon },g\left( \epsilon \right) \right) $, where $\epsilon $ is a
perturbation parameter, and $M_{\epsilon }$ is a manifold associated with $%
\epsilon $, with a metric $g\left( \epsilon \right) $ defined on it, such
that $\left( M_{\epsilon },g\left( \epsilon \right) \right) $ tends to $%
\left( M_{\mathrm{o}},\tensor[^{\left(0\right)}]{g}{}\right) $ as $\epsilon
\rightarrow 0$. Then, a perturbation $\delta X$ of the quantity $X$ is
defined as the pull-back of $X$ from $M_{\epsilon }$ to $M_{\mathrm{o}}$
through a map between the two manifolds, subtracted by $X$ on $M_{\mathrm{o}%
} $. By Taylor expansion, $\delta X$ can be split into pieces of orders of $%
\epsilon $; i.e., $\delta X=\tensor[^{\left(1\right)}]{X}{}+%
\tensor[^{\left(2\right)}]{X}{}+O\left( \epsilon ^{3}\right) $, where $%
\tensor[^{\left(1\right)}]{X}{}$ and $\tensor[^{\left(2\right)}]{X}{}$
denote first-order and second-order perturbations, respectively.

A gauge transformation associated with a perturbation corresponds to
changing a map between $M_{\mathrm{o}}$ and $M_{\epsilon }$. Then, a gauge
transformation of a first-order perturbation is described by $%
\tensor[^{\left(1\right)}]{X}{}^{\prime }-\tensor[^{\left(1\right)}]{X}{}=%
\mathcal{L}_{\xi }X$, where $\tensor[^{\left(1\right)}]{X}{}$ and $%
\tensor[^{\left(1\right)}]{X}{}^{\prime }$ are first-order perturbations via
two different maps, and $\mathcal{L}_{\xi }X$ denotes the Lie derivative of $%
X$ with respect to $\xi $, a vector of $O\left( \epsilon \right) $ defined
from the two maps. A first-order perturbation $\tensor[^{\left(1%
\right)}]{X}{}$ is guage-invariant if $\mathcal{L}_{\xi }X$ vanishes for all 
$\xi $: it is possible only if $X$ is zero, or a constant scalar, or
constructed by Kronecker delta with constant coefficients. This approach was
first introduced in \cite{stewart1974} and reviewed later in \cite%
{stewart1990}.

Let us consider the Minkowski spacetime as the background manifold $M_{%
\mathrm{o}}$. Then, a first-order perturbation of the Riemann tensor is
gauge-invariant as the Riemann tensor vanishes in the background spacetime.
With the stress-energy tensor being assumed to vanish to first order in $%
\epsilon $, a first-order perturbation of Einstein equations becomes the
classical gravitational wave equations: $\partial ^{c}\partial _{c}h_{ab}=0$%
, with gauge conditions $\partial ^{b}h_{ab}=0$ and $\tensor{h}{^{a}_{a}}=0$%
, where $h_{ab}$ denotes a first-order perturbation of a metric $g_{ab}$,
and $\partial $ is the partial derivative associated with the metric $%
\tensor[^{\left(0\right)}]{g}{}$ in the Minkowski spacetime. We consider a
monochromatic plane wave solution: $h_{ab}=\mathcal{H}_{ab}e^{\mathrm{i}P}$,
where $\mathcal{H}_{ab}$ is a complex constant amplitude, and $P$ is a real
phase scalar that satisfies $\tensor[^{\left(0\right)}]{n}{^{a}}\partial
_{a}P<0$ for an arbitrary observer with the 4-velocity $\tensor[^{\left(0%
\right)}]{n}{^{a}}$ in the Minkowski spacetime and vanishes at the second
covariant derivative. Also, the propagation vector for gravitational waves
is defined as $k^{a}\equiv \partial ^{a}P$, which is a constant null vector.

For an inertial observer with $\tensor[^{\left(0\right)}]{n}{^{a}}$ living
in the Minkowski spacetime, the propagation vector $k^{a}$ can be decomposed
into $k^{a}=\omega _{\mathrm{g}}\left( \tensor[^{\left(0\right)}]{n}{^{a}}%
+\kappa ^{a}\right) $, where $\omega _{\mathrm{g}}\equiv -%
\tensor[^{\left(0\right)}]{n}{^{a}}\partial _{a}P>0$ is the frequency of
gravitational waves measured by the observer, and $\kappa ^{a}$ is a spatial
unit vector orthogonal to $\tensor[^{\left(0\right)}]{n}{^{a}}$ in the
observer's point of view. Further, if we impose an additional gauge
condition $h_{ab}\tensor[^{\left(0\right)}]{n}{^{b}}=0$ (which together with 
$\tensor{h}{^{a}_{a}}=0$ constitutes a transverse-traceless gauge), then the
\textquotedblleft electric\textquotedblright\ part of the Riemann tensor
perturbation becomes $\tensor[^{\left(1\right)}]{R}{_{acbd}}%
\tensor[^{\left(0\right)}]{n}{^{c}}\tensor[^{\left(0\right)}]{n}{^{d}}=\frac{%
1}{2}\omega _{\mathrm{g}}^{2}\tensor{h}{_{ab}}$: it should be noted that $%
h_{ab}$ in our gauge conditions is proportional to a gauge-invariant
perturbation $\tensor[^{\left(1\right)}]{R}{_{acbd}}\tensor[^{\left(0%
\right)}]{n}{^{c}}\tensor[^{\left(0\right)}]{n}{^{d}}$.

Although the observer's 4-velocity $n^{a}$ is not constant in a perturbed
spacetime $M_{\epsilon }$, we can impose the geodesic condition $n^{b}\nabla
_{b}n^{a}=0$, with $\nabla $ being the covariant derivative associated with
the metric $g\left( \epsilon \right) $ in $M_{\epsilon }$, which reduces to $%
\tensor[^{\left(0\right)}]{n}{^{b}}\partial _{b}\tensor[^{\left(1%
\right)}]{n}{^{a}}=0$ in $M_{\mathrm{o}}$; i.e., $\tensor[^{\left(1%
\right)}]{n}{^{a}}$ can be set to a constant along $\tensor[^{\left(0%
\right)}]{n}{^{a}}$. For instance, we can set $\tensor[^{\left(1%
\right)}]{n}{^{a}}=0$, which implies that the 4-velocity of the geodesic
observer in $M_{\epsilon }$ is not perturbed by gravitational waves to
first-order; namely, $\delta n^{a}=O\left( \epsilon ^{2}\right) $.
Throughout our analysis, we set $\tensor[^{\left(1\right)}]{n}{^{a}}=0$, and 
$\delta n^{a}=0$ to first order in $\epsilon $.

Taking account of Einstein-Maxwell equations, we assume no electric charge
for a source. Through the equations, the influence of the electromagnetic
field on the spacetime geometry is of second order, and a first-order
perturbation of the equations becomes the same classical gravitational wave
equations as above.

In the Lorenz gauge $\partial ^{a}\tensor[^{\left(0\right)}]{A}{_{a}}=0$,
Maxwell equations become the classical electromagnetic wave equations: $%
\partial ^{b}\partial _{b}\tensor[^{\left(0\right)}]{A}{_{a}}=0$, where $%
\tensor[^{\left(0\right)}]{A}{_{a}}$ denotes an electromagnetic potential in 
$M_{\mathrm{o}}$. Again, we consider a monochromatic plane wave solution: $%
\tensor[^{\left(0\right)}]{A}{_{a}}=\tensor[^{\left(0\right)}]{%
\mathcal{A}}{_{a}}e^{\mathrm{j}\tensor[^{\left(0\right)}]{Q}{}}$, where $%
\tensor[^{\left(0\right)}]{\mathcal{A}}{_{a}}$ is a complex constant
amplitude, and $\tensor[^{\left(0\right)}]{Q}{}$ is a real phase scalar that
satisfies $\tensor[^{\left(0\right)}]{n}{^{a}}\partial _{a}%
\tensor[^{\left(0\right)}]{Q}{}<0$ for an arbitrary observer with the
4-velocity $\tensor[^{\left(0\right)}]{n}{^{a}}$ in the Minkowski spacetime\
and vanishes at the second covariant derivative. Also, the propagation
vector for electromagnetic waves is defined as $\tensor[^{\left(0%
\right)}]{l}{^{a}}\equiv \partial ^{a}\tensor[^{\left(0\right)}]{Q}{}$,
which is a constant null vector. Note that one must distinguish the complex
numbers $\mathrm{i}$ and $\mathrm{j}$ assigned to describe gravitational
waves and electromagnetic waves, respectively: in particular, when they are
mixed in quadratic forms, such as $\mathrm{ij}$ or $\mathrm{ji}$. As $%
\mathrm{i}$ and $\mathrm{j}$ describe a phase shift by $\pi /2$ with respect
to the reference phase of each independent wave, they must be treated
separately. We impose the commutativity of multiplication between $\mathrm{i}
$ and $\mathrm{j}$; namely, $\mathrm{ij=}$ $\mathrm{ji}$.

For an inertial observer with $\tensor[^{\left(0\right)}]{n}{^{a}}$ living
in the Minkowski spacetime, the propagation vector $\tensor[^{\left(0%
\right)}]{l}{^{a}}$ can be decomposed into $\tensor[^{\left(0%
\right)}]{l}{^{a}}=\tensor[^{\left(0\right)}]{\omega}{_{\mathrm{e}}}\left( %
\tensor[^{\left(0\right)}]{n}{^{a}}+\tensor[^{\left(0\right)}]{\lambda}{^{a}}%
\right) $, where $\tensor[^{\left(0\right)}]{\omega}{_{\mathrm{e}}}=-%
\tensor[^{\left(0\right)}]{n}{^{a}}\partial_{a}\tensor[^{\left(0%
\right)}]{Q}{}>0$ is the frequency of electromagnetic waves measured by the
observer, and $\tensor[^{\left(0\right)}]{\lambda}{^{a}}$ is a spatial unit
vector orthogonal to $\tensor[^{\left(0\right)}]{n}{^{a}}$ in the observer's
point of view. Further, we impose an additional gauge condition $%
\tensor[^{\left(0\right)}]{A}{_{a}}\tensor[^{\left(0\right)}]{n}{^{a}}=0$.
Then, electric and magnetic fields become $\tensor[^{\left(0%
\right)}]{E}{_{a}}=\mathrm{j}\tensor[^{\left(0\right)}]{\omega}{_{%
\mathrm{e}}}\tensor[^{\left(0\right)}]{A}{_{a}}$ and $\tensor[^{\left(0%
\right)}]{B}{_{a}}=\mathrm{j}\tensor[^{\left(0\right)}]{\omega}{_{%
\mathrm{e}}}\tensor[^{\left(0\right)}]{\epsilon}{_{ab}^{c}}%
\tensor[^{\left(0\right)}]{\lambda}{^{b}}\tensor[^{\left(0\right)}]{A}{_{c}}$%
, respectively: it should be noted that $\tensor[^{\left(0\right)}]{A}{_{a}}$
in our gauge conditions is proportional to a gauge-invariant quantity $%
\tensor[^{\left(0\right)}]{E}{_{a}}$ or $\tensor[^{\left(0\right)}]{B}{_{a}}$%
.

As shown in \cref{app:adapted_frame_ef}, we can find a right-handed
orthonormal frame $\left\{ \tensor[^{\left(0\right)}]{n}{^{a}},%
\tensor[^{\left(0\right)}]{p}{^{a}},\tensor[^{\left(0\right)}]{s}{^{a}},%
\tensor[^{\left(0\right)}]{\lambda}{^{a}}\right\} $, in which the electric
field is written in the form $\tensor[^{\left(0\right)}]{E}{_{a}}=\left( %
\tensor[^{\left(0\right)}]{\mathcal{E}}{_{\left(p\right)}}%
\tensor[^{\left(0\right)}]{p}{_{a}}+\mathrm{j}\tensor[^{\left(0\right)}]{%
\mathcal{E}}{_{\left(s\right)}}\tensor[^{\left(0\right)}]{s}{_{a}}\right) e^{%
\mathrm{j}\tensor[^{\left(0\right)}]{Q}{}}$, where $\tensor[^{\left(0%
\right)}]{\mathcal{E}}{_{\left(p\right)}}$ and $\tensor[^{\left(0\right)}]{%
\mathcal{E}}{_{\left(s\right)}}$ are real scalars, representing the axes of
polarization ellipse. The ellipticity of polarization is defined as $%
\tensor[^{\left(0\right)}]{\chi}{}\equiv \tan ^{-1}\left( %
\tensor[^{\left(0\right)}]{\mathcal{E}}{_{\left(s\right)}}/%
\tensor[^{\left(0\right)}]{\mathcal{E}}{_{\left(p\right)}}\right) $.

The intensity of electromagnetic waves on the plane orthogonal to a unit
vector $z^{a}$ is obtained from the time average of a Poynting vector $%
\tensor[^{\left(0\right)}]{P}{^{a}}=\tensor[^{\left(0\right)}]{%
\epsilon}{^{abc}}\Re\left(\tensor[^{\left(0\right)}]{E}{_{b}}\right)\Re\left(%
\tensor[^{\left(0\right)}]{B}{_{c}}\right)$ contracted with $z^{a}$; i.e., $%
\mathrm{Intensity}=\left\langle z^{a}\tensor[^{\left(0\right)}]{P}{_{a}}%
\right\rangle $. The time average of a quantity $f\left( t\right) $ is
defined as $\left\langle f\left(t\right)\right\rangle
=\lim_{T\rightarrow\infty}\frac{1}{T}\int_{t}^{t+T}f\left( t^{\prime
}\right) dt^{\prime }$, where $t$ is the proper time measured by an observer
with $\tensor[^{\left(0\right)}]{n}{^{a}}$ \cite{reitz_foundations_2009}.

Stokes parameters describe the polarization state of light that are obtained
by measurements of intensities using optical devices, e.g., polarizers and
waveplates: $\tensor[^{\left(0\right)}]{S}{_{0}}$ is a measure of the total
intensity of light, $\tensor[^{\left(0\right)}]{S}{_{1}}$ and $%
\tensor[^{\left(0\right)}]{S}{_{2}}$ jointly describe the linear
polarization, and $\tensor[^{\left(0\right)}]{S}{_{3}}$ describes the
circular polarization. In the adapted frame $\left\{\tensor[^{\left(0%
\right)}]{n}{^{a}},\tensor[^{\left(0\right)}]{p}{^{a}},\tensor[^{\left(0%
\right)}]{s}{^{a}},\tensor[^{\left(0\right)}]{\lambda}{^{a}}\right\} $ for
the electric field $\tensor[^{\left(0\right)}]{E}{_{a}}=\left( %
\tensor[^{\left(0\right)}]{\mathcal{E}}{_{\left(p\right)}}%
\tensor[^{\left(0\right)}]{p}{_{a}}+\mathrm{j}\tensor[^{\left(0\right)}]{%
\mathcal{E}}{_{\left(s\right)}}\tensor[^{\left(0\right)}]{s}{_{a}}\right) e^{%
\mathrm{j}\tensor[^{\left(0\right)}]{Q}{}}$, we obtain $\tensor[^{\left(0%
\right)}]{S}{_{0}}=\frac{1}{2}\left(\tensor[^{\left(0\right)}]{%
\mathcal{E}}{^{2}_{\left(p\right)}}+\tensor[^{\left(0\right)}]{%
\mathcal{E}}{^{2}_{\left(s\right)}}\right) $, $\tensor[^{\left(0%
\right)}]{S}{_{1}}=\frac{1}{2}\left( \tensor[^{\left(0\right)}]{%
\mathcal{E}}{^{2}_{\left(p\right)}}-\tensor[^{\left(0\right)}]{%
\mathcal{E}}{^{2}_{\left(s\right)}}\right) $, $\tensor[^{\left(0%
\right)}]{S}{_{2}}=0$, and $\tensor[^{\left(0\right)}]{S}{_{3}}=%
\tensor[^{\left(0\right)}]{\mathcal{E}}{_{\left(p\right)}}%
\tensor[^{\left(0\right)}]{\mathcal{E}}{_{\left(s\right)}}$.

\section{Perturbation of Electromagnetic Waves}

\label{sec:perturbation_of_emw}

Let us consider a perturbed spacetime $M_{\epsilon }$ in which the Riemann
tensor does not vanish. Then Maxwell equations are written as: 
\begin{equation}
\nabla ^{b}\nabla _{b}A_{a}=\tensor{R}{^{b}_{a}}A_{b},  \label{eq:me}
\end{equation}%
with the Lorenz gauge condition, 
\begin{equation}
\nabla ^{a}A_{a}=0,  \label{eq:lg}
\end{equation}%
where $R_{ab}=\tensor{R}{^{c}_{acb}}$ denotes the Ricci tensor. \cref{eq:me}
presents inhomogeneous Maxwell equations, namely,\ electromagnetic wave
equations extended to the curved (perturbed) spacetime $M_{\epsilon }$. Then
we can write down a solution in the form: 
\begin{equation}
A_{a}=\mathcal{A}_{a}e^{\mathrm{j}Q},  \label{eq:A}
\end{equation}%
where $\mathcal{A}_{a}$ and $Q$ correspond to the amplitude and phase of an
electromagnetic wave, respectively in the Minkowski spacetime $M_{\mathrm{o}%
} $. Note that $\mathcal{A}_{a}$ is not constant in general, unlike its
counterpart in $M_{\mathrm{o}}$, and that $l^{a}\equiv \nabla ^{a}Q$ is not
null in general, unlike its counterpart in $M_{\mathrm{o}}$. Then for the
violation of the null condition, we define a quantity: 
\begin{equation}
\nu \equiv g_{ab}l^{a}l^{b}.  \label{eq:nu}
\end{equation}%
Also, $A_{a}$ is not a transverse wave due to $A_{a}l^{a}\neq 0$. Following
from \cref{eq:A}, the field strength tensor $F_{ab}=\nabla _{a}A_{b}-\nabla
_{b}A_{a}$ is expressed in the same form: 
\begin{equation}
F_{ab}=\mathcal{F}_{ab}e^{\mathrm{j}Q},  \label{eq:Fab}
\end{equation}%
where 
\begin{equation}
\mathcal{F}_{ab}\equiv 2\nabla _{\lbrack a}\mathcal{A}_{b]}+2\mathrm{j}l_{[a}%
\mathcal{A}_{b]}.  \label{eq:F_tilde}
\end{equation}

Let us consider a geodesic observer with $n^{a}$ in $M_{\epsilon }$, as
mentioned in the previous section. We introduce the 3+1 formalism to split a
tensor into temporal and spatial parts, using a projection tensor $\gamma
_{ab}$ defined as 
\begin{equation}
\tensor{\gamma}{^{a}_{b}}\equiv \tensor{\delta}{^{a}_{b}}+n^{a}n_{b}.
\end{equation}%
Then the propagation vector $l^{a}$ can be decomposed into the temporal
component $\omega _{\mathrm{e}}$ and the spatial unit vector $\lambda ^{a}$,
given respectively by 
\begin{align}
\omega _{\mathrm{e}}& \equiv -n^{a}l_{a},  \label{eq:omega_l} \\
\lambda ^{a}& \equiv \frac{\tensor{\gamma}{^{a}_{b}}l^{b}}{\sqrt{\gamma
_{cd}l^{c}l^{d}}}.  \label{eq:lambda}
\end{align}%
The electric and magnetic fields as measured by an observer with $n^{a}$ can
be expressed in the same form as \cref{eq:A}: 
\begin{align}
E_{a}& =\mathcal{E}_{a}e^{\mathrm{j}Q},  \label{eq:E} \\
B_{a}& =\mathcal{B}_{a}e^{\mathrm{j}Q},  \label{eq;B}
\end{align}%
where%
\begin{align}
\mathcal{E}_{a}& =\mathcal{F}_{ab}n^{b},  \label{eq:E_tilde} \\
\mathcal{B}_{a}& =\frac{1}{2}\tensor{\epsilon}{^{bc}_{a}}\mathcal{F}_{bc},
\label{eq:B_tilde}
\end{align}%
where $\mathcal{F}_{ab}$ refers to \cref{eq:F_tilde}, and $\epsilon
_{abc}\equiv n^{d}\varepsilon _{dabc}$ denotes the `spatial' Levi-Civita
tensor while $\varepsilon _{dabc}$ is the `spacetime' Levi-Civita tensor.

According to \cref{app:adapted_frame_ef}, we can introduce an adapted
orthonormal frame $\left\{ p^{a},s^{a}\right\} $ into $M_{\epsilon }$ with
no restriction, in which one can express 
\begin{equation}
\mathcal{E}^{a}=\mathcal{E}_{\left( p\right) }p^{a}+\mathrm{j}\mathcal{E}%
_{\left( s\right) }s^{a},  \label{eq:E_a}
\end{equation}%
for the electric field \cref{eq:E}, where the axes of polarization ellipse, $%
\mathcal{E}_{\left( p\right) }$ and $\mathcal{E}_{\left( s\right) }$ are
real scalars given by 
\begin{align}
\mathcal{E}_{\left( p\right) }& \equiv \sqrt{\Re \left( \mathcal{E}%
_{a}\right) \Re \left( \mathcal{E}_{b}\right) g^{ab}},  \label{eq:E_p} \\
\mathcal{E}_{\left( s\right) }& \equiv \sqrt{\Im \left( \mathcal{E}%
_{a}\right) \Im \left( \mathcal{E}_{b}\right) g^{ab}},  \label{eq:E_s}
\end{align}%
where $\Re \left( {}\right) $ and $\Im \left( {}\right) $ are defined by 
\begin{align}
\Re (f)& =\frac{1}{2}\left( f+f^{\ast }\right) , \\
\Im (f)& =\frac{1}{2\mathrm{j}}\left( f-f^{\ast }\right) ,
\end{align}%
where $f$ is a complex quantity and $\ast $ denotes the complex conjugate
with respect to $\mathrm{j}$. There are multiple possible pairs of $\left\{
p^{a},s^{a}\right\} $ in $M_{\epsilon }$, but we choose one, whose values in
the limit $\epsilon \rightarrow 0$ coincide with $\left\{
p^{a},s^{a}\right\} $ in $M_{\mathrm{o}}$. Also, we define the ellipticity
as 
\begin{equation}
\chi \equiv \tan ^{-1}\left( \frac{\mathcal{E}_{\left( s\right) }}{\mathcal{E%
}_{\left( p\right) }}\right) .  \label{eq:chi}
\end{equation}%
Note that the polarization plane spanned by $\left\{ p^{a},s^{a}\right\} $
is not orthogonal to $\lambda ^{a}$ in general. Then for the violation of
the transversity condition, we define a quantity: 
\begin{equation}
\tau \equiv \lambda ^{a}E_{a}.  \label{eq:tau}
\end{equation}

Now, let us consider a first-order perturbation of Maxwell equations. From %
\cref{eq:me} it becomes 
\begin{equation}
\partial ^{b}\partial _{b}\tensor[^{\left(1\right)}]{A}{_{a}}=2%
\tensor{C}{^{bc}_{a}}\partial _{c}\tensor[^{\left(0\right)}]{A}{_{b}}%
+h^{bc}\partial _{b}\partial _{c}\tensor[^{\left(0\right)}]{A}{_{a}},
\label{eq:pme}
\end{equation}%
where 
\begin{align}
\tensor{C}{^{a}_{bc}}& \equiv \tensor{\mathcal{C}}{^{a}_{bc}}e^{\mathrm{i}P},
\label{eq:C} \\
\tensor{\mathcal{C}}{^{a}_{bc}}& \equiv \frac{\mathrm{i}}{2}\left( %
\tensor{\mathcal{H}}{^{a}_{b}}k_{c}+\tensor{\mathcal{H}}{^{a}_{c}}k_{b}-%
\mathcal{H}_{bc}k^{a}\right) .  \label{eq:C1}
\end{align}%
Note that the perturbation of the Ricci tensor term from the right-hand side
of \cref{eq:me} vanishes: it is due to the transverse-traceless gauge.
Assuming a solution in the form $\tensor[^{\left(1\right)}]{A}{_{a}}\sim 
\mathrm{const.}\times e^{\mathrm{i}P}e^{\mathrm{j}\tensor[^{\left(0%
\right)}]{Q}{}}$, it is obtained as: 
\begin{align}
\tensor[^{\left(1\right)}]{A}{_{a}}& =\left( -\mathrm{i}\frac{%
\tensor{\mathcal{C}}{^{b}_{ca}}\tensor[^{\left(0\right)}]{\mathcal{A}}{_{b}}%
\tensor[^{\left(0\right)}]{l}{^{c}}}{k^{d}\tensor[^{\left(0\right)}]{l}{_{d}}%
}-\mathrm{i}\mathrm{j}\frac{\mathcal{H}_{bc}\tensor[^{\left(0%
\right)}]{l}{^{b}}\tensor[^{\left(0\right)}]{l}{^{c}}\tensor[^{\left(0%
\right)}]{\mathcal{A}}{_{a}}}{2k^{d}\tensor[^{\left(0\right)}]{l}{_{d}}}%
\right)  \nonumber \\
& \qquad \times e^{\mathrm{i}P}e^{\mathrm{j}\tensor[^{\left(0\right)}]{Q}{}}.
\label{eq:A_1_sol}
\end{align}%
Here we rule out the case of $k^{a}\tensor[^{\left(0\right)}]{l}{_{a}}=0$,
in which the right-hand side of \cref{eq:pme} vanishes and hence
gravitational waves do not affect the electromagnetic wave to the
first-order perturbation.

From \cref{eq:A} perturbation of $A_{a}$ to first order yields 
\begin{equation}
\tensor[^{\left(1\right)}]{A}{_{a}}=\left( \tensor[^{\left(1\right)}]{%
\mathcal{A}}{_{a}}+\mathrm{j}\tensor[^{\left(0\right)}]{\mathcal{A}}{_{a}}%
\tensor[^{\left(1\right)}]{Q}{}\right) e^{\mathrm{j}\tensor[^{\left(0%
\right)}]{Q}{}}.  \label{eq:Aa1}
\end{equation}%
Matching this with \cref{eq:A_1_sol}, on the right-hand sides of the two
equations, the first terms correspond to each other and so do the second
terms: as $l^{a}\propto \tensor[^{\left(0\right)}]{\omega}{_{\mathrm{e}}}$
(from \cref{eq:omega_l}), the first terms are at $O\left( %
\tensor[^{\left(0\right)}]{\omega}{}_{\mathrm{e}}^{0}\right) $ while the
second terms are at $O\left( \tensor[^{\left(0\right)}]{\omega}{}_{\mathrm{e}%
}^{1}\right) $, where $\tensor[^{\left(0\right)}]{\omega}{_{\mathrm{e}}}$
serves as an order-counting parameter in the \textit{geometrical-optics}
approach \cite{dolan_geometrical_2018}. Then the perturbations of amplitude
and phase, $\tensor[^{\left(1\right)}]{\mathcal{A}}{}$ and $%
\tensor[^{\left(1\right)}]{Q}{}$ can be identified respectively as: 
\begin{align}
\tensor[^{\left(1\right)}]{Q}{}& =-\mathrm{i}\frac{\mathcal{H}_{bc}%
\tensor[^{\left(0\right)}]{l}{^{b}}\tensor[^{\left(0\right)}]{l}{^{c}}}{%
2k^{d}\tensor[^{\left(0\right)}]{l}{_{d}}}e^{\mathrm{i}P},
\label{eq:Q_perturb} \\
\tensor[^{\left(1\right)}]{\mathcal{A}}{_{a}}& =-\mathrm{i}\frac{%
\tensor{\mathcal{C}}{^{b}_{ca}}\tensor[^{\left(0\right)}]{\mathcal{A}}{_{b}}%
\tensor[^{\left(0\right)}]{l}{^{c}}}{k^{d}\tensor[^{\left(0\right)}]{l}{_{d}}%
}e^{\mathrm{i}P}.  \label{eq:A_perturb}
\end{align}%
Here one should be careful about the denominators, expressed by $\sim k^{a}%
\tensor[^{\left(0\right)}]{l}{_{a}}=\omega _{\mathrm{g}}\tensor[^{\left(0%
\right)}]{\omega}{_{\mathrm{e}}}\left( -1+\cos \theta \right) $, where $%
\theta $ is the angle between the propagation directions of\ gravitational
and electromagnetic waves. As $\theta \rightarrow 0$, the first term inside
the round brackets of \cref{eq:A_1_sol} diverges while the second term
converges; hence, separation of the terms by order of $\tensor[^{\left(0%
\right)}]{\omega}{_{\mathrm{e}}}$ fails. Then the value of $\theta $ for the
geometrical-optics approach to be valid is given by 
\begin{equation}
\theta \gtrsim \sqrt{2\frac{\omega _{\mathrm{g}}}{\tensor[^{\left(0%
\right)}]{\omega}{_{\mathrm{e}}}}}.  \label{eq:theta}
\end{equation}

Substituting \cref{eq:Q_perturb} into the first-order perturbation of $\nu $
from \cref{eq:nu} through $l^{a}=\nabla ^{a}Q$, we obtain 
\begin{equation}
\tensor[^{\left(1\right)}]{\nu}{}=0.  \label{eq:nu1}
\end{equation}%
This means that the null condition for the electromagnetic wave is not
violated to the first-order perturbation. In the same manner, from
perturbation of \cref{eq:omega_l}, we obtain the fractional perturbation of
the electromagnetic wave frequency, given by 
\begin{equation}
\frac{\tensor[^{\left(1\right)}]{\omega}{_{\mathrm{e}}}}{\tensor[^{\left(0%
\right)}]{\omega}{_{\mathrm{e}}}}=-\frac{1}{2\Theta }\mathcal{H}_{ab}%
\tensor[^{\left(0\right)}]{\lambda}{^{a}}\tensor[^{\left(0\right)}]{%
\lambda}{^{b}}e^{\mathrm{i}P},  \label{eq:omega_frac}
\end{equation}%
where $\Theta \equiv 1-\kappa ^{a}\tensor[^{\left(0\right)}]{\lambda}{_{a}}%
=1-\cos \theta $. Note that $\tensor[^{\left(1\right)}]{\nu}{}$ and $%
\tensor[^{\left(1\right)}]{\omega}{_{\mathrm{e}}}$ are gauge-invariant as
they are constant scalars evaluated in $M_{\mathrm{o}}$.

From \cref{eq:F_tilde} perturbation of $\mathcal{F}_{ab}$ to first order
yields 
\begin{equation}
\tensor[^{\left(1\right)}]{\mathcal{F}}{_{ab}}=2\partial _{\lbrack a}%
\tensor[^{\left(1\right)}]{\mathcal{A}}{_{b]}}+2\mathrm{j}\left( \partial
_{\lbrack a}\tensor[^{\left(1\right)}]{Q}{}\right) \tensor[^{\left(0%
\right)}]{\mathcal{A}}{_{b]}}+2\mathrm{j}\tensor[^{\left(0\right)}]{l}{_{[a}}%
\tensor[^{\left(1\right)}]{\mathcal{A}}{_{b]}}.  \label{eq:F_perturb}
\end{equation}%
Here the first term on the right-hand side, being at $O\left(\omega_{\mathrm{%
g}}\tensor[^{\left(0\right)}]{\omega}{}_{\mathrm{e}}^{0}\right) $, can be
ignored in comparison with the other terms at $O\left(\tensor[^{\left(0%
\right)}]{\omega}{}_{\mathrm{e}}^{1}\right)$ in the geometrical-optics
approximation, provided that $\tensor[^{\left(0\right)}]{\omega}{}^{1}_{%
\mathrm{e}}\gg\omega_{\mathrm{g}}\tensor[^{\left(0\right)}]{\omega}{}^{0}_{%
\mathrm{e}}$. Now, from \cref{eq:E_tilde,eq:B_tilde} the first-order
perturbations of $\mathcal{E}_{a}$ and $\mathcal{B}_{a}$ are given by 
\begin{align}
\tensor[^{\left(1\right)}]{\mathcal{E}}{_{a}}& =\tensor[^{\left(1\right)}]{
\mathcal{F}}{_{ab}}\tensor[^{\left(0\right)}]{n}{^{b}},
\label{eq:E_cal_perturb} \\
\tensor[^{\left(1\right)}]{\mathcal{B}}{_{a}}& =\frac{1}{2}%
\tensor[^{\left(0\right)}]{\epsilon}{^{bc}_{a}}\tensor[^{\left(1\right)}]{%
\mathcal{F}}{_{bc}}-\tensor[^{\left(0\right)}]{\epsilon}{^{bc}_{a}}%
\tensor{h}{^{d}_{b}}\tensor[^{\left(0\right)}]{\mathcal{F}}{_{dc}}.
\label{eq:B_cal_perturb}
\end{align}%
In \cref{eq:B_cal_perturb} we have used the fact that the perturbation of
the spatial Levi-Civita tensor vanishes: 
\begin{equation}
\tensor[^{\left(1\right)}]{\epsilon}{_{abc}}=0,  \label{eq:slc}
\end{equation}%
which is shown in \cref{app:perturbation_of_Levi-Civita_tensor}.

Using these together with %
\cref{eq:C1,eq:Q_perturb,eq:A_perturb,eq:F_perturb,eq:E_cal_perturb} for the
perturbations of $\mathcal{E}_{\left( p\right) }$ and $\mathcal{E}_{\left(
s\right) }$ from \cref{eq:E_p,eq:E_s}, we obtain the fractional
perturbations of the axes of polarization ellipse, given by 
\begin{equation}
\frac{\tensor[^{\left( 1\right) }]{\mathcal{E}}{_{\left( p\right)}}}{%
\tensor[^{\left(0\right)}]{\mathcal{E}}{_{\left(p\right)}}}=\frac{%
\tensor[^{\left( 1\right)}]{\mathcal{E}}{_{\left( s\right)}}}{%
\tensor[^{\left(0\right)}]{\mathcal{E}}{_{\left(s\right)}}}=-\frac{1}{%
2\Theta }\mathcal{H}_{ab}\tensor[^{\left(0\right)}]{\lambda}{^{a}}%
\tensor[^{\left(0\right)}]{\lambda}{^{b}}e^{\mathrm{i}P}.  \label{eq:E_frac}
\end{equation}%
Due to this, however, the perturbation of the ellipticity $\chi $ from %
\cref{eq:chi} vanishes: 
\begin{equation}
\tensor[^{\left(1\right)}]{\chi}{}=0.  \label{eq:chi_perturb}
\end{equation}%
This means that the ellipticity is maintained while the perturbed axes of
polarization ellipse oscillate with gravitational waves: they expand and
shrink periodically together.

Using \cref{eq:lambda,eq:E,eq:E_tilde,eq:E_cal_perturb} for \cref{eq:tau},
we obtain the perturbation of $\tau $: 
\begin{equation}
\tensor[^{\left(1\right)}]{\tau}{}=0.  \label{eq:tau_perturb}
\end{equation}%
This means that the transversity condition for the electromagnetic wave is
not violated to the first-order perturbation. Note that $%
\tensor[^{\left(
1\right) }]{\mathcal{E}}{_{\left( p\right)}}$, $%
\tensor[^{\left(
1\right)}]{\mathcal{E}}{_{\left( s\right)}}$, $\tensor[^{\left(1\right)}]{%
\chi}{}$, and $\tensor[^{\left(1\right)}]{\tau}{}$ are gauge-invariant as
they are constant scalars evaluated in $M_{\mathrm{o}}$.

\section{Application: Observation of Gravitational Waves}

\label{sec:application}

With regard to the observation of gravitational waves, one can consider the
measurement of Stokes parameters and their perturbations. This requires that
the Stokes parameters be measured in a perturbed spacetime $M_{\epsilon }$.
For this purpose, we introduce another right-handed orthonormal frame $%
\left\{ n^{a},x^{a},y^{a},z^{a}\right\} $, where $z^{a}$ is directed along
the time-averaged Poynting vector $P^{a}$ for electromagnetic waves, given
by 
\begin{equation}
z^{a}\equiv \frac{\left\langle P^{a}\right\rangle }{\sqrt{g_{bc}\left\langle
P^{b}\right\rangle \left\langle P^{c}\right\rangle }}.  \label{eq:z}
\end{equation}
Here the Poynting vector $P^{a}$ is defined as the cross product of the real
electric and magnetic fields in $M_{\epsilon }$: 
\begin{align}
P^{a}&=\tensor{\epsilon}{^{abc}}\Re\left(E_{b}\right)\Re\left(B_{c}\right).
\end{align}
And the time average of a quantity $f\left( t\right) $ is defined as 
\begin{align}
\left\langle f\left(t\right)\right\rangle =\frac{1}{T}\int_{t}^{t+T}f\left(
t^{\prime }\right) dt^{\prime },
\end{align}
where $t$ is the proper time measured by an observer with $n^{a}$, and $T$
is a time scale such that $\tensor{\omega}{_{\mathrm{e}}}T\gg 1 \gg%
\tensor{\omega}{_{\mathrm{g}}}T$ with $T$ covering a finite number of
oscillation periods of the electromagnetic fields; therefore, defined
through the time average, $z^{a}$ contains the perturbation oscillating at $%
\tensor{\omega}{_{\mathrm{g}}}$, exhibiting the effects of gravitational
waves, while its unperturbed part is constant. $x^{a}$ and $y^{a}$ are
perturbed accordingly while being orthogonal to $z^{a}$ and to each other.
We consider that the frame $\left\{ n^{a},x^{a},y^{a},z^{a}\right\} $
defined in this manner is experimentally feasible as it is determined
naturally based on an observable; namely, the time-averaged Poynting vector.

In this frame, the Stokes parameters are expressed as: 
\begin{equation}
4\pi S_{I}=\frac{1}{2}\left\langle \Re \left\{ \mathfrak{p}_{I}^{ab}\left( 
\mathcal{E}_{a}\mathcal{B}_{b}^{\ast }+\mathcal{E}_{a}\mathcal{B}_{b}e^{2%
\mathrm{j}Q}\right) \right\} \right\rangle ,  \label{eq:Stokes_parameters}
\end{equation}%
where $I=0,1,2,3$, 
and $\mathfrak{p}_{I}^{ab}$ denote projections defined by 
\begin{align}
\mathfrak{p}_{0}^{ab}& \equiv x^{a}y^{b}-y^{a}x^{b},  \label{eq:P0} \\
\mathfrak{p}_{1}^{ab}& \equiv x^{a}y^{b}+y^{a}x^{b},  \label{eq:P1} \\
\mathfrak{p}_{2}^{ab}& \equiv -x^{a}x^{b}+y^{a}y^{b},  \label{eq:P2} \\
\mathfrak{p}_{3}^{ab}& \equiv -\mathrm{j}\left( x^{a}x^{b}+y^{a}y^{b}\right)
.  \label{eq:P3}
\end{align}%
It should be noted that the Stokes parameters $S_{I}$ in $M_{\epsilon }$,
being defined through the time average as above, contain the perturbations
oscillating at $\tensor{\omega}{_{\mathrm{g}}}$, induced by gravitational
waves, unlike their counterparts in $M_{\mathrm{o}}$. Using rotational
properties of $S_{1}$ and $S_{2}$, we set $\left\{ x^{a},y^{a}\right\} $
such that Stokes parameter $S_{2}$ vanishes. There are two possible sets of $%
\left\{ n^{a},x^{a},y^{a},z^{a}\right\} $ in $M_{\epsilon }$ satisfying
this, but we choose one, whose values in the limit $\epsilon \rightarrow 0$
coincide with $\left\{ \tensor[^{\left(0\right)}]{n}{^{a}},%
\tensor[^{\left(0\right)}]{p}{^{a}},\tensor[^{\left(0\right)}]{s}{^{a}},%
\tensor[^{\left(0\right)}]{\lambda}{^{a}}\right\} $ in $M_{\mathrm{o}}$.

Now, we consider perturbations of the Stokes parameters. For this sake, we
need a perturbation of the spatial Levi-Civita tensor, as given by %
\cref{eq:slc}, together with the components along $\tensor[^{\left(0%
\right)}]{p}{^{a}}$ and $\tensor[^{\left(0\right)}]{s}{^{a}}$ of the
perturbations of the frame $\left\{ x^{a},y^{a}\right\} $, 
\begin{align}
\tensor[^{\left(1\right)}]{x}{^{a}}\tensor[^{\left(0\right)}]{p}{_{a}}& =-%
\frac{1}{2}\mathcal{H}_{ab}\tensor[^{\left(0\right)}]{p}{^{a}}%
\tensor[^{\left(0\right)}]{p}{^{b}}e^{\mathrm{i}P},  \label{eq:x_comp} \\
\tensor[^{\left(1\right)}]{y}{^{a}}\tensor[^{\left(0\right)}]{s}{_{a}}& =-%
\frac{1}{2}\mathcal{H}_{ab}\tensor[^{\left(0\right)}]{s}{^{a}}%
\tensor[^{\left(0\right)}]{s}{^{b}}e^{\mathrm{i}P},  \label{eq:y_comp}
\end{align}%
the derivations of which are given in \cref{app:perturbation_of_frame}.
Using these, we obtain the fractional perturbations of the Stokes parameters
from \cref{eq:Stokes_parameters}, given by 
\begin{align}
\frac{\tensor[^{\left(1\right)}]{S}{_{0}}}{\tensor[^{\left(0%
\right)}]{S}{_{0}}}& =-\frac{1}{\Theta }\mathcal{H}_{ab}\tensor[^{\left(0%
\right)}]{\lambda}{^{a}}\tensor[^{\left(0\right)}]{\lambda}{^{b}}e^{\mathrm{i%
}P},  \label{eq:fractional_change_S0} \\
\frac{\tensor[^{\left(1\right)}]{S}{_{1}}}{\tensor[^{\left(0%
\right)}]{S}{_{1}}}& =-\frac{1}{\Theta }\mathcal{H}_{ab}\tensor[^{\left(0%
\right)}]{\lambda}{^{a}}\tensor[^{\left(0\right)}]{\lambda}{^{b}}e^{\mathrm{i%
}P},  \label{eq:fractional_change_S1} \\
\tensor[^{\left(1\right)}]{S}{_{2}}& =0,  \label{eq:fractional_change_S2} \\
\frac{\tensor[^{\left(1\right)}]{S}{_{3}}}{\tensor[^{\left(0%
\right)}]{S}{_{3}}}& =-\frac{1}{\Theta }\mathcal{H}_{ab}\tensor[^{\left(0%
\right)}]{\lambda}{^{a}}\tensor[^{\left(0\right)}]{\lambda}{^{b}}e^{\mathrm{i%
}P}.  \label{eq:fractional_change_S3}
\end{align}%
Note that these are gauge-invariant quantities as they are constant scalars
evaluated in $M_{\mathrm{o}}$.

In relation to the observation of gravitational waves, let us discuss the
antenna patterns for the measurement of $h$ through the Stokes parameters,
defined by 
\begin{equation}
h\equiv -\frac{1}{\Theta }\mathcal{H}_{ab}\tensor[^{\left(0\right)}]{%
\lambda}{^{a}}\tensor[^{\left(0\right)}]{\lambda}{^{b}}e^{\mathrm{i}P}.
\label{eq:measure}
\end{equation}%
Now, we introduce an adapted frame for gravitational waves. As shown in %
\cref{app:adapted_frame_gw}, one can find a right-handed orthonormal frame $%
\left\{ \tensor[^{\left(0\right)}]{n}{^{a}},u^{a},v^{a},\kappa ^{a}\right\} $%
, in which the amplitude of gravitational waves is expressed as: 
\begin{equation}
\mathcal{H}_{ab}=\mathcal{H}_{+}e_{ab}^{+}+\mathrm{i}\mathcal{H}_{\times
}e_{ab}^{\times },  \label{eq:H_cal}
\end{equation}%
where $\mathcal{H}_{+}$ and $\mathcal{H}_{\times }$ are real scalars, the
projections of the amplitude onto the polarization tensors $e_{ab}^{+}$ and $%
e_{ab}^{\times }$, respectively, defined by 
\begin{align}
e_{ab}^{+}& \equiv u_{a}u_{b}-v_{a}v_{b}, \\
e_{ab}^{\times }& \equiv u_{a}v_{b}+v_{a}u_{b}.
\end{align}%
Then the Euler angles $\left( \phi ,\theta ,\psi \right) $ as defined in
Ref. \cite{goldstein_classical_2002,pai_data-analysis_2001} yield the
following relations between the adapted \textit{spatial }frames $\left\{ %
\tensor[^{\left(0\right)}]{p}{^{a}},\tensor[^{\left(0\right)}]{s}{^{a}},%
\tensor[^{\left(0\right)}]{\lambda}{^{a}}\right\} $ and $\left\{
u^{a},v^{a},\kappa ^{a}\right\} $ for electromagnetic and gravitational
waves, respectively: 
\begin{align}
u^{a}& =R\left( \tensor[^{\left(0\right)}]{\lambda}{^{a}},\phi \right)
R\left( \tensor[^{\left(0\right)}]{s}{^{a}},\theta \right) R\left( %
\tensor[^{\left(0\right)}]{\lambda}{^{a}},\psi \right) \tensor[^{\left(0%
\right)}]{p}{^{a}}, \\
v^{a}& =R\left( \tensor[^{\left(0\right)}]{\lambda}{^{a}},\phi \right)
R\left( \tensor[^{\left(0\right)}]{s}{^{a}},\theta \right) R\left( %
\tensor[^{\left(0\right)}]{\lambda}{^{a}},\psi \right) \tensor[^{\left(0%
\right)}]{s}{^{a}}, \\
\kappa ^{a}& =R\left( \tensor[^{\left(0\right)}]{\lambda}{^{a}},\phi \right)
R\left( \tensor[^{\left(0\right)}]{s}{^{a}},\theta \right) R\left( %
\tensor[^{\left(0\right)}]{\lambda}{^{a}},\psi \right) \tensor[^{\left(0%
\right)}]{\lambda}{^{a}},
\end{align}%
where $R\left( \beta ^{a},\alpha \right) $ denotes a rotation of a vector by
the angle $\alpha $ with respect to the axis $\beta ^{a}$.

\begin{figure}[tbp]
\includegraphics[width=0.45\textwidth]{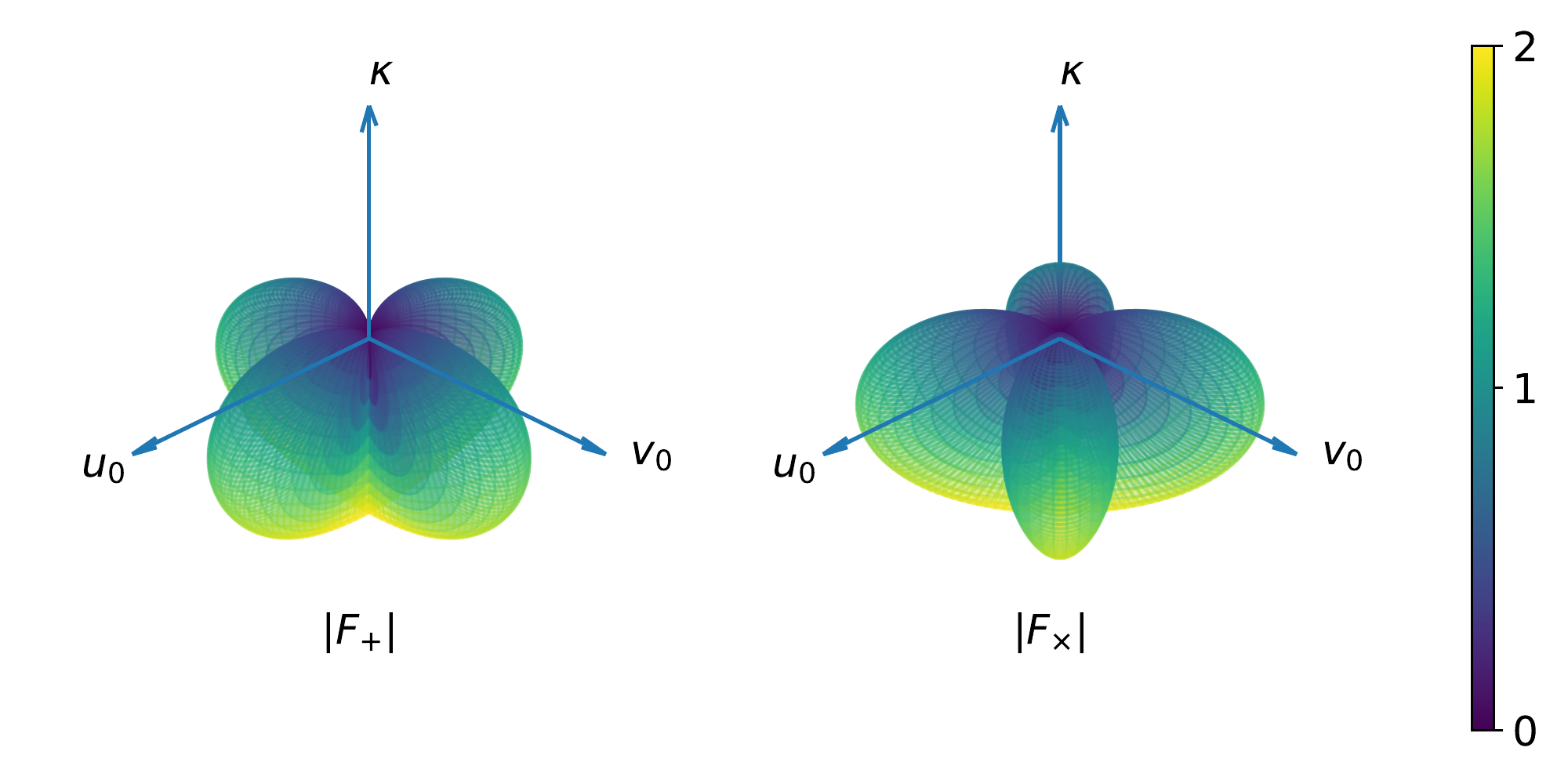} \centering
\caption{Antenna patterns of $\left\vert F_{+}\right\vert $ and $\left\vert
F_{\times }\right\vert $ for the measurement of $h$; as viewed in
consideration with gravitational waves propagating along the direction of
the\ $\boldsymbol{\protect\kappa }$-axis while being polarized in the $%
\boldsymbol{uv}$-plane in a quadrupole manner. The frame $\left\{ 
\boldsymbol{u}_{0},\boldsymbol{v}_{0},\boldsymbol{\protect\kappa }\right\} $
is defined from an adapted frame $\left\{ \boldsymbol{u},\boldsymbol{v},%
\boldsymbol{\protect\kappa }\right\} $ with $\protect\psi =0$. In this
frame, the polar angle is defined as $\protect\vartheta =\protect\pi -%
\protect\theta $, where $\protect\theta$ refers to the angle between the
propagation directions of gravitational and electromagnetic waves, whereas
the azimuthal angle is defined as $\protect\varphi=\protect\pi/2+\protect%
\psi $. }
\label{fig:antenna}
\end{figure}

Substituting \cref{eq:H_cal} into \cref{eq:measure}, we obtain 
\begin{equation}
h=\left( \mathcal{H}_{+}F_{+}+\mathrm{i}\mathcal{H}_{\times }F_{\times
}\right) e^{iP},
\end{equation}%
where $F_{+}$ and $F_{\times }$ are the antenna patterns for $+$ and $\times 
$ polarization states, respectively, given by 
\begin{align}
F_{+}& =-2\cos ^{2}\left( \theta /2\right) \cos \left( 2\psi \right) ,
\label{eq:Fplus} \\
F_{\times }& =2\cos ^{2}\left( \theta /2\right) \sin \left( 2\psi \right) .
\label{eq:Fcross}
\end{align}%
These are in agreement with Refs. \cite%
{chamberlin_stochastic_2012,alves_pulsar_2011,lee_pulsar_2008,yunes_gravitational-wave_2013}%
, although our angles $\theta $ and $\psi $ are defined in a slightly
different manner than the ones in the literature. In \cref{fig:antenna} are
presented the antenna patterns of $\left\vert F_{+}\right\vert $ and $%
\left\vert F_{\times }\right\vert $ as viewed in consideration with
gravitational waves, which propagate along the direction of the\ $%
\boldsymbol{\kappa }$-axis while being polarized in the $\boldsymbol{uv}$%
-plane in a quadrupole manner. We evaluate the \textit{sky} \textit{averages}
of the following quantities over $\left( \phi ,\theta ,\psi \right) $: 
\begin{align}
\left\langle F_{+}F_{+}\right\rangle & =\frac{2}{3}, \\
\left\langle F_{\times }F_{\times }\right\rangle & =\frac{2}{3}, \\
\left\langle F_{+}F_{\times }\right\rangle & =0.
\end{align}%
Then the angular efficiency factor as defined in Ref. \cite{Maggiore2007} is 
\begin{equation}
F\equiv \left\langle F_{+}F_{+}\right\rangle +\left\langle F_{\times
}F_{\times }\right\rangle =\frac{4}{3}.  \label{eq:F}
\end{equation}%
And the sky average of $\left\vert h\right\vert ^{2}$ is given by 
\begin{equation}
\left\langle \left\vert h\right\vert ^{2}\right\rangle =\frac{2}{3}\left( 
\mathcal{H}_{+}^{2}+\mathcal{H}_{\times }^{2}\right) .  \label{eq:h_sq}
\end{equation}

\section{Discussion}

We have worked out a perturbation of electromagnetic waves due to
gravitational waves from the perturbed Maxwell equations, as given by %
\crefrange{eq:Aa1}{eq:A_perturb}. The perturbation has two oscillatory parts
from electromagnetic and gravitational waves. Inspecting it closely, the
amplitude of the perturbed electromagnetic waves in general contains the
perturbations of amplitude and phase mixed together in it. Then to
disentangle the mixed perturbations from each other, we invoke the
geometrical-optics approach, adopting the frequency of electromagnetic waves
as an order-counting parameter. However, it should be noted that the
geometrical-optics approach can break down if the propagation directions of
electromagnetic and gravitational waves are extremely close to each other;
so close as to almost coincide, as shown by our analysis with \cref{eq:theta}%
.

From this perturbation analysis, we have found that the null condition for
the electromagnetic wave is not violated to the first-order perturbation, as
shown by \cref{eq:nu1}. Also, we have obtained the fractional perturbation
of \ the electromagnetic wave frequency, as given by \cref{eq:omega_frac},
and confirmed that it is equivalent to the gravitational-wave-induced
redshift in the literature \cite{detweiler_pulsar_1979}. In addition, the
axes of polarization ellipse defined via the perturbed electric field
exhibit the oscillatory feature of gravitational waves, as shown by %
\cref{eq:E_frac}: they expand and shrink periodically together. However, the
ellipticity and the orthogonality between the polarization ellipse and the
propagation direction of the electromagnetic wave are preserved to the
first-order perturbation due to gravitational waves, as evidenced by %
\cref{eq:chi_perturb,eq:tau_perturb}.

We have employed Stokes parameters as optical observables containing the
gravitational wave signals. The measurement of the Stokes parameters
requires a suitable orthonormal frame in which the parameters are expressed.
To this end, we set one spatial unit vector to be along the propagation
direction of the time-averaged Poynting vector, and arrange the other two
vectors perpendicular to this such that the Stokes parameter $S_{2}$
vanishes. We have obtained the fractional perturbations of the other
non-vanishing Stokes parameters, as given by %
\cref{eq:fractional_change_S0,eq:fractional_change_S1,eq:fractional_change_S3}%
, which have turned out to be all identical. Interestingly, apart from the
factor $2$, these are also identical to the fractional perturbations of the
electromagnetic wave frequency and the axes of polarization ellipse, as
given by \cref{eq:omega_frac,eq:E_frac}. The antenna patterns defined
through the Stokes parameters, as given by \cref{eq:Fplus,eq:Fcross}, show
that the detectability of gravitational wave signals vanishes when the
propagation directions of gravitational and electromagnetic waves are
opposite to each other, whereas the detectability becomes the maximum when
the two waves propagate in the same direction, as described in %
\cref{fig:antenna}.

Our analysis might find its application to some practical issue in relation
to Cosmic Microwave Background (CMB) anisotropies, characterized by scalar
perturbations due to temperature fluctuations and tensor perturbations due
to generation of polarization. As for the latter, free charges, being
agitated by primordial gravitational waves propagating through the CMB
plasma, rescatter electromagnetic radiation, and this imprints a
characteristic pattern of linear polarization on the CMB map, represented by
E-modes and B-modes, which are derived by means of the Stokes parameters 
\cite%
{kovac_detection_2002,hu_cmb_1997,hu_cmb_1997-1,kamionkowski_statistics_1997,zaldarriaga_all-sky_1997,kim_how_2011,kim_primordial_2016}%
. However, if other gravitational waves from different sources pass through
while observing the CMB radiation, the Stokes parameters will be perturbed
as given by %
\cref{eq:fractional_change_S0,eq:fractional_change_S1,eq:fractional_change_S3}%
, causing the polarization pattern to change through the perturbed E-modes
and B-modes accordingly. Therefore, for accurate measurement of the effects
from primordial gravitational waves alone, the effects from other
gravitational waves should be identified and disentangled carefully, based
on our analysis of polarization.

In this work, we have investigated the effect of gravitational waves on one
light ray. But it will also be fairly interesting to study the effect of
gravitational waves in a situation where two or more light rays interfere
with each other. With respect to this, most analyses regarding the
interferometry in the literature have concentrated on the phase perturbation
so far. However, a more complete analysis of the interferometry would
require consideration of the effect from the amplitude perturbation as well.
In addition, it will be interesting to study the effect of gravitational
waves in a situation where the polarization directions of two or more light
rays are not aligned. We leave discussion of all these for the future
research.

\begin{acknowledgments}
C. Park appreciates Gungwon Kang for valuable discussions. C. Park was supported in part by the Basic Science Research Program through the National Research Foundation of Korea (NRF) funded by the Ministry of Education (NRF-2018R1D1A1B07041004), and by the National Institute for Mathematical Sciences (NIMS) funded by Ministry of Science and ICT (B20710000). 
D.-H. Kim was supported by the Basic Science Research Program through the National Research Foundation of Korea (NRF) funded by the Ministry of Education (NRF-2018R1D1A1B07051276).
\end{acknowledgments}

\appendix

\section{Adapted Frame for Electric Field}

\label{app:adapted_frame_ef}

Given an electric field $E_{a}=\mathcal{E}_{a}e^{\mathrm{j}Q}$, one can
consider that there always exists a suitable phase $\alpha $ to make 
\begin{equation}
E_{a}=\mathcal{E}_{a}e^{-\mathrm{j}\alpha }e^{\mathrm{j}\left( Q+\alpha
\right) }=\mathcal{E}_{a}^{\prime }e^{\mathrm{j}Q^{\prime }},  \label{A1}
\end{equation}%
such that 
\begin{equation}
\Re _{\mathrm{j}}\left( \mathcal{E}_{a}^{\prime }\right) \Im _{\mathrm{j}%
}\left( \mathcal{E}_{b}^{\prime }\right) g^{ab}=0,  \label{A2}
\end{equation}%
where $\Re _{\mathrm{j}}\left( {}\right) $ and $\Im _{\mathrm{j}}\left(
{}\right) $ are defined by 
\begin{align}
\Re _{\mathrm{j}}(f)& =\frac{1}{2}\left( f+f^{\ast }\right) , \\
\Im _{\mathrm{j}}(f)& =\frac{1}{2\mathrm{j}}\left( f-f^{\ast }\right) ,
\end{align}%
where $f$ is a complex quantity and $\ast $ denotes the complex conjugate
with respect to $\mathrm{j}$. Dropping the sign $^{\prime }$ from the
left-hand side, Eq. (\ref{A2}) is the orthogonality condition for the
decomposed parts of $\mathcal{E}^{a}$; namely, $\Re _{\mathrm{j}}\left( 
\mathcal{E}^{a}\right) \perp \Im _{\mathrm{j}}\left( \mathcal{E}^{a}\right) $%
. Out of this, we can construct an orthonormal frame $\left\{
p^{a},s^{a}\right\} $ by 
\begin{align}
p^{a}& \equiv \frac{\Re _{\mathrm{j}}\left( \mathcal{E}^{a}\right) }{%
\left\Vert \Re _{\mathrm{j}}\left( \mathcal{E}\right) \right\Vert }, \\
s^{a}& \equiv \frac{\Im _{\mathrm{j}}\left( \mathcal{E}^{a}\right) }{%
\left\Vert \Im _{\mathrm{j}}\left( \mathcal{E}\right) \right\Vert },
\end{align}%
where $\left\Vert X\right\Vert \equiv \sqrt{X^{a}X^{b}g_{ab}}$ for a vector $%
X^{a}$. In this frame, $\mathcal{E}^{a}$ is expressed in the form: 
\begin{equation}
\mathcal{E}^{a}=\mathcal{E}_{\left( p\right) }p^{a}+\mathrm{j}\mathcal{E}%
_{\left( s\right) }s^{a},  \label{eq:E_in_natural_frame}
\end{equation}%
where $\mathcal{E}_{\left( p\right) }\equiv \left\Vert \Re _{\mathrm{j}%
}\left( \mathcal{E}\right) \right\Vert $ and $\mathcal{E}_{\left( s\right)
}\equiv \left\Vert \Im _{\mathrm{j}}\left( \mathcal{E}\right) \right\Vert $
are real scalars.

\section{Adapted Frame for Gravitational Waves}

\label{app:adapted_frame_gw}

Given a gravitational wave $h_{ab}=\mathcal{H}_{ab}e^{\mathrm{i}P}$, one can
consider that there always exists a suitable phase $\beta $ to make%
\begin{equation}
h_{ab}=\mathcal{H}_{ab}e^{-\mathrm{i}\beta }e^{\mathrm{i}\left( P+\beta
\right) }=\mathcal{H}_{ab}^{\prime }e^{\mathrm{i}P^{\prime }},  \label{B1}
\end{equation}%
such that 
\begin{equation}
\Re _{\mathrm{i}}\left( \mathcal{H}_{ab}^{\prime }\right) \Im _{\mathrm{i}%
}\left( \mathcal{H}_{cd}^{\prime }\right) g^{ac}g^{bd}=0,  \label{B2}
\end{equation}%
where $\Re _{\mathrm{i}}\left( {}\right) $ and $\Im _{\mathrm{i}}\left(
{}\right) $ are defined by 
\begin{align}
\Re _{\mathrm{i}}(f)& =\frac{1}{2}\left( f+f^{\ast }\right) , \\
\Im _{\mathrm{i}}(f)& =\frac{1}{2\mathrm{i}}\left( f-f^{\ast }\right) ,
\end{align}%
where $f$ is a complex quantity and $\ast $ denotes the complex conjugate
with respect to $\mathrm{i}$. Dropping the sign $^{\prime }$ from the
left-hand side, Eq. (\ref{B2}) is the orthogonality condition for the
decomposed parts of $\mathcal{H}_{ab}$; namely, $\Re _{\mathrm{i}}\left( 
\mathcal{H}_{ab}\right) \perp \Im _{\mathrm{i}}\left( \mathcal{H}%
_{ab}\right) $. Out of this, we can construct a tensor basis $\left\{
e_{ab}^{+},e_{ab}^{\times }\right\} $ by 
\begin{align}
e_{ab}^{+}& =\sqrt{2}\frac{\Re _{\mathrm{i}}\left( \mathcal{H}_{ab}\right) }{%
\left\Vert \Re _{\mathrm{i}}\left( \mathcal{H}\right) \right\Vert },
\label{B4} \\
e_{ab}^{\times }& =\sqrt{2}\frac{\Im _{\mathrm{i}}\left( \mathcal{H}%
_{ab}\right) }{\left\Vert \Im _{\mathrm{i}}\left( \mathcal{H}\right)
\right\Vert },  \label{B5}
\end{align}%
where $\left\Vert T\right\Vert \equiv \sqrt{T_{ab}T_{cd}g^{ac}g^{bd}}$ for a
rank $\left( 0,2\right) $ tensor $T_{ab}$. In this basis, $\mathcal{H}_{ab}$
is expressed in the form: 
\begin{equation}
\mathcal{H}_{ab}=\mathcal{H}_{+}e_{ab}^{+}+\mathrm{i}\mathcal{H}_{\times
}e_{ab}^{\times },  \label{B6}
\end{equation}%
where $\mathcal{H}_{+}\equiv \left\Vert \Re _{\mathrm{i}}\left( \mathcal{H}%
\right) \right\Vert /\sqrt{2}$ and $\mathcal{H}_{\times }\equiv \left\Vert
\Im _{\mathrm{i}}\left( \mathcal{H}\right) \right\Vert /\sqrt{2}$ are real
scalars.{}

We can create an orthonormal frame $\left\{ u^{a},v^{a}\right\} $ such that
linear combinations of $u^{a}$ and $v^{a}$ satisfy the following equalities
in view of Eqs. (\ref{B4}) to (\ref{B6}):%
\begin{eqnarray}
\Re _{\mathrm{i}}\left( \mathcal{H}_{ab}\right) \left( c_{1}u^{a}+c_{2}%
\mathrm{i}v^{a}\right) &=&c_{3}u_{b}+c_{4}\mathrm{i}v_{b},  \label{B7} \\
\Im _{\mathrm{i}}\left( \mathcal{H}_{ab}\right) \left( c_{1}u^{a}+c_{2}%
\mathrm{i}v^{a}\right) &=&c_{5}u_{b}+c_{6}\mathrm{i}v_{b},  \label{B8}
\end{eqnarray}%
where $c_{1}$, $c_{2}$, $c_{3}$, $c_{4}$, $c_{5}$, $c_{6}$ are undetermined
coefficients. Contracting both sides of Eqs. (\ref{B7}) and (\ref{B8}) with $%
c_{1}u^{b}+c_{2}\mathrm{i}v^{b}$, and using $u^{b}u_{b}=v^{b}v_{b}=1$ and $%
u^{b}v_{b}=0$, we find 
\begin{align}
\Re _{\mathrm{i}}\left( \mathcal{H}_{ab}\right) \left[ \left(
u^{a}u^{b}-v^{a}v^{b}\right) +\mathrm{i}\left( u^{a}v^{a}+v^{a}u^{a}\right) %
\right] =c_{3}-c_{4},&  \label{B9} \\
\Im _{\mathrm{i}}\left( \mathcal{H}_{ab}\right) \left[ \left(
u^{a}u^{b}-v^{a}v^{b}\right) +\mathrm{i}\left( u^{a}v^{a}+v^{a}u^{a}\right) %
\right] =c_{5}-c_{6},&  \label{B10}
\end{align}%
where setting $c_{1}=c_{2}$ has simplified the equalities. Now, defining the
tensor basis by 
\begin{align}
e_{ab}^{+}& \equiv u_{a}u_{b}-v_{a}v_{b},  \label{B11} \\
e_{ab}^{\times }& \equiv u_{a}v_{b}+v_{a}u_{b},  \label{B12}
\end{align}%
and using Eqs. (\ref{B4}) to (\ref{B6}) for Eqs. (\ref{B9}) and (\ref{B10}),
we determine $c_{1}=c_{2}=1$, $c_{3}=-c_{4}=\mathcal{H}_{+}$, $c_{5}=-c_{6}=%
\mathrm{i}\mathcal{H}_{\times }$. Then from Eqs. (\ref{B9}) and (\ref{B10})
we establish%
\begin{equation}
\mathcal{H}_{ab}\left( e^{+ab}+\mathrm{i}e^{\times ab}\right) =2\left( 
\mathcal{H}_{+}+\mathrm{i}\mathcal{H}_{\times }\right) ,
\end{equation}%
which is equivalent to Eq. (\ref{B6}).

\section{Perturbation of Levi-Civita Tensor}

\label{app:perturbation_of_Levi-Civita_tensor}

The spacetime Levi-Civita tensor $\varepsilon _{abcd}$ is normalized by 
\begin{equation}
-4!=\varepsilon _{abcd}\varepsilon _{efgh}g^{ae}g^{bf}g^{cg}g^{dh}.
\end{equation}%
From this perturbation of the tensor to first order yields 
\begin{equation}
\tensor[^{\left(1\right)}]{\varepsilon}{_{abcd}}=\frac{1}{2}%
\tensor{h}{^{e}_{e}}\tensor[^{\left(0\right)}]{\varepsilon}{_{abcd}}.
\end{equation}%
By our choice of the perturbation gauge, $\tensor{h}{^{a}_{a}}=0$, $%
\tensor[^{\left(1\right)}]{\varepsilon}{_{abcd}}$ vanishes. Then this leads
to the perturbation of the spatial Levi-Civita tensor $\epsilon _{abc}\equiv
n^{d}\varepsilon _{dabc}$ (for a geodesic observer with the 4-velocity $%
n^{a} $) vanishing too: 
\begin{equation}
\tensor[^{\left(1\right)}]{\epsilon}{_{abc}}=\tensor[^{\left(0%
\right)}]{n}{^{d}}\tensor[^{\left(1\right)}]{\varepsilon}{_{dabc}}+%
\tensor[^{\left(0\right)}]{\varepsilon}{_{dabc}}\tensor[^{\left(1%
\right)}]{n}{^{d}}=0.
\end{equation}

\section{Perturbation of Normalized Frame}

\label{app:perturbation_of_frame}

The orthonormal frame $\left\{ x^{a},y^{a}\right\} $ for the observation of
Stokes parameters must satisfy 
\begin{align}
x^{a}x^{b}g_{ab}& =y^{a}y^{b}g_{ab}=1,  \label{D1} \\
x^{a}y^{b}g_{ab}& =0.
\end{align}%
With no perturbation, the frame $\left\{ x^{a},y^{a}\right\} $ would tend to 
$\left\{ \tensor[^{\left(0\right)}]{p}{^{a}},\tensor[^{\left(0%
\right)}]{s}{^{a}}\right\} $ in the background. Then from the normalization
condition (\ref{D1}), perturbation of $x^{a}$ and $y^{a}$ to first order
yields 
\begin{align}
\tensor[^{\left(1\right)}]{x}{^{a}}\tensor[^{\left(0\right)}]{p}{_{a}}& =-%
\frac{1}{2}h_{ab}\tensor[^{\left(0\right)}]{p}{^{a}}\tensor[^{\left(0%
\right)}]{p}{^{b}}, \\
\tensor[^{\left(1\right)}]{y}{^{a}}\tensor[^{\left(0\right)}]{s}{_{a}}& =-%
\frac{1}{2}h_{ab}\tensor[^{\left(0\right)}]{s}{^{a}}\tensor[^{\left(0%
\right)}]{s}{^{b}}.
\end{align}

\bibliographystyle{unsrt}
\bibliography{references}

\end{document}